\documentclass[amssymb,twocolumn,10pt,prl]{revtex4}
\usepackage{dcolumn}
\usepackage{bm}

\begin{document} 
\noindent
{\large \bf Comment on ``Bell's Theorem without Inequalities and without 
Alignments"} \\

In this Comment we show that Cabello's proof of Bell's theorem
 without inequalities~\cite{cab} does not exhibit two of the three 
 {\em ``remarkable properties"} which the proof is claimed
 to possess.
More precisely it is our purpose to show that property (c), 
 stating that {\em ``local 
 observables can be measured by means of tests on individual qubits"}
 is definitely false since it has been obtained by a wrong use of the
 basic rules of the quantum formalism, while property (b), stating
 that {\em ``distant local setups do not need to be aligned since the required
 perfect correlations are achieved for any local rotation of the 
 setups"}, is also wrong due to the failure of validity of property (c). 
    
Let us first review the argument devised by the author in support of 
 property (c). 
In the Letter~\cite{cab} it is claimed that, in order to measure 
 the observable $F$ of Eq.(4) whose outcomes, given the state 
 $\vert \eta \rangle$ of Eq.(1), can be equal either to $+1$ or $-1$, 
 it is enough to perform individual measurements of the set of 
 spin-observables ($\sigma_{z1}$,$\sigma_{z2}$,$\sigma_{x3}$,$\sigma_{x4}$) 
 onto the four Alice's particles. 
In fact, according to the author, after having performed such set of
 measurements, it is sufficient to check if the (unique) eigenstate associated 
 to one of the 16 possible sets of outcomes appears in the expansion of 
 $\vert \phi_{0}\rangle$ of Eq.(23) or of $\vert \phi_{1}\rangle$ of Eq.(24). 
In the first case the observable $F$ is given the value $-1$, while in 
 the second case the value $+1$ is attributed.
Of course this implies that, according to the basic postulates of quantum 
 mechanics, such attributed values must be observed with 
 certainty if a measurement of $F$ is actually performed instantaneously 
 after the sequence of single-particle spin measurements.

We will now show that this argument supporting property (c) is totally 
 incorrect by resorting to a simple and clarifying example~\footnote{An
 equivalent remark can be raised about the measurement 
 process of the observable $G$ of Eq.~(5).}.
In fact, let us suppose we have obtained, in the individual 
 spin-measurements, all the spins up in the considered directions 
 (our conclusion holding true also for all other possible sets of outcomes). 
In this case we must conclude, as prescribed in the Letter, that after 
 such measurements the value $+1$ should be attributed to the observable $F$
 since the eigenstate $\vert 00\bar{0}\bar{0}\rangle$ associated to the
 above-mentioned set of outcomes appears within $\vert \phi_{1}\rangle$.
However, as a consequence of the Reduction Postulate, the initial state 
 $\vert \eta \rangle$ collapses to $\vert \tilde{\eta} \rangle=
 1/2\,\vert 00\bar{0}\bar{0}\rangle\otimes (\vert \psi_{0} \rangle+
 \sqrt{3}\vert \psi_{1}\rangle)$ instantaneously after the
 spin-measurements.
Now in this collapsed state, since $\vert \langle  00\bar{0}\bar{0} 
 \vert \phi_{1} \rangle \vert^{2}= 1/12$, the probability that Alice's 
 measurement of observable $F$ gives the result $+1$ is clearly different 
 from $1$~\footnote{Moreover, given $\vert \tilde{\eta} \rangle$, no certain
 prediction concerning Bob's measurement of observable $G$ can be made too.}.  
This is obviously due to the fact that, in the new state $\vert \tilde{\eta}
 \rangle$, there is a non-zero probability of obtaining the neglected outcome 
 $F=0$, still holding true that the other measurement outcome $F=-1$ cannot be
 obtained.
The fact that before Alice's spin-measurements the only possible outcomes 
 of $F$ were $\pm 1$, does not affect our conclusion since, in that situation, 
 the state of the system $\vert \eta \rangle$ was different from the one
 $\vert \tilde{\eta} \rangle$ which results after the spin-measurements.\\
The situation would have been different if the observable $F$ had been
 a function (in the Dirac sense) of the set of observables 
 ($\sigma_{z1},\sigma_{z2},\sigma_{x3},\sigma_{x4}$). 
In this case in fact, it would have been possible to attribute a definite value
 to $F$ by the knowledge of the outcomes of the single-particle
 spin-measurements. 
So, concluding this first remark, it is clearly not possible to assign a
 definite value to the observable $F$ by performing the set
 of local spin-measurements described in~\cite{cab}: the only way to do that
 is to make actually a (four-qubit) collective measurement of $F$, thus
 forcing 
 Alice's state to collapse onto $\vert \phi_{0} \rangle$ or $\vert \phi_{1}
 \rangle$. \\
Let us now pass to review property (b). According to the author, Alice's  
 setups for measuring observable $F$ do not need to be aligned
 with those located at Bob's place, since {\em ``perfect correlations
 are achieved for any local rotation of the local setups"}.
Though, as a consequence of the failure of validity of property (c), the 
 outcomes of the observable $F$ (as well as those of $G$) must be obtained by 
 performing a complicated collective measurement involving four
 particles (instead of the four distinct single-particle spin-measurements 
 along two arbitrary orthogonal directions, as described in the Letter).
Therefore the invariance property of $F$ and $G$ under the restricted class 
 of tensor products of four equal single-qubit rotations, cannot be used to 
 conclude that the desired perfect correlations between their outcomes
 are achieved {\em ``for any local rotation of the local setups"} simply
 because $F$ and $G$ cannot be measured by means of single-qubit 
 spin-measurements.
Therefore Cabello's argument, as it has been formulated, still requires 
 {\em perfect correlations} 
 and {\em perfect alignments} between the involved observables and measuring 
 setups respectively, like all known proofs of Bell's theorem.  
 
The author thanks Dr. Danilo Mauro for his valuable comments.  
   
\noindent Luca Marinatto [marinatto@ts.infn.it]\\
 Department of Theoretical Physics, University of Trieste, Trieste and
 Istituto Nazionale di Fisica Nucleare, Sez. di Trieste, Italy.

\end{document}